\documentclass[9pt,final]{IEEEtran}
\usepackage{graphicx,amsmath,amssymb,array}
\usepackage[noadjust]{cite}
\newtheorem{theorem}{\textbf{Theorem}}
\newtheorem{definition}{\textbf{Definition}}
\usepackage{times}
\usepackage{epsf}
\usepackage{epsfig}
\usepackage{amsmath}
\usepackage{amsfonts}
\usepackage{amssymb}
\usepackage{amstext}
\usepackage{latexsym}
\usepackage{color}
\usepackage{ifthen}
\usepackage{multirow}
\usepackage{subfigure}

\newcommand{\be}{\begin{equation}}
\newcommand{\ee}{\end{equation}}
\newcommand{\bear}{\begin{eqnarray}}
\newcommand{\eear}{\end{eqnarray}}
\newcommand{\bears}{\begin{eqnarray*}}
\newcommand{\eears}{\end{eqnarray*}}
\newcommand{\bi}{\begin{itemize}}
\newcommand{\ei}{\end{itemize}}
\newcommand{\ben}{\begin{enumerate}}
\newcommand{\een}{\end{enumerate}}

\newcommand{\Expt}{\mbox{${\mathbb E}$} }

\renewcommand{\vec}[1]{\mbox{\boldmath$#1$}}

\renewcommand{\vec}[1]{\mbox{\boldmath$#1$}}

\begin{document}

\title{Multiuser Successive Refinement and Multiple Description Coding}
\author{Chao Tian,~\IEEEmembership{Member,~IEEE}, Jun Chen,~\IEEEmembership{Member,~IEEE}, Suhas N. Diggavi,~\IEEEmembership{Member,~IEEE}
\thanks{This work was presented in part at 
2006 IEEE Information Theory Workshop in Oct. 2006, Chengdu, China.}}
\date{}

\maketitle

\maketitle
\begin{abstract}
We consider the multiuser successive refinement (MSR) problem, where the users are connected to a central server via links with different noiseless capacities, and each user wishes to reconstruct in a successive-refinement fashion. An achievable region is given for the two-user two-layer case and it provides the complete rate-distortion region for the Gaussian source under the MSE distortion measure. The key observation is that this problem includes the multiple description (MD) problem (with two descriptions) as a subsystem, and the techniques useful in the MD problem can be extended to this case. It is shown that the coding scheme based on the universality of random binning is sub-optimal, because multiple Gaussian side informations only at the decoders do incur performance loss, in contrast to the case of single side information at the decoder. 
It is further shown that unlike the single user case, when there are multiple users, the loss of performance by a multistage coding approach can be unbounded for the Gaussian source.
The result suggests that in such a setting, the benefit of using successive refinement is not likely to justify the accompanying performance loss. The MSR problem is also related to the source coding problem where each decoder has its individual side information, while the encoder has the complete set of the side informations. The MSR problem further includes several variations of the MD problem, for which the specialization of the general result is investigated and the implication is discussed.   
\end{abstract}

\section{Introduction}
\label{sec:intro}
Multiuser information theory has attracted much attention recently because of the growth in the complexity and capability of the practical communication networks. In this work, we consider the multiuser successive refinement (MSR) problem formulated by Pradhan and Ramchandran in \cite{Pradhanramchandran:02}. In this problem, a server is to provide multimedia data to users connected to the server through channels with different (noiseless) capacities, e.g., a dial-up connection vs. a high-speed cable connection. The server performs the transmission in a broadcasting manner in order to reduce operating cost, and thus the users with bad channels will only receive a (known) subset of the bitstream, while the users with good channels will be able to receive the complete bitstream. Furthermore, to reduce the delay for each user, the server would also like to provide the bitstream in a successive refinement fashion user-wise. The ``multiusers" in the MSR problem thus receive degraded message sets, while the ``successive refinement" refers to the fact that there are multiple rounds (layers) of such transmissions. 

\begin{figure}[tb]
  \centering 
\includegraphics[scale=0.42]{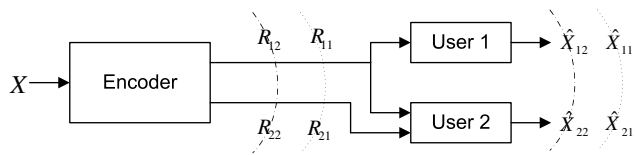}
\caption{A system diagram with two users and two layers. \label{fig:systemdiag} The ``good" user (user 2) receives the complete message while the ``bad" user (user 1) only receives a subset of the message. There are two rounds of transmissions, illustrated by the dotted curve and dash-dot curve, and in each round the users form their reconstructions accordingly.}
\end{figure}

A diagram is given in  Fig. \ref{fig:systemdiag} for a system with two users and two layers. We will assume the user with good channel connection will remain so for the complete transmission, however the exact rates $R_{11}$ (the first subscript specifies which user and the second subscript specifies which round of transmission), $R_{12},R_{21},R_{22}$ can vary. If the transmission rate is fixed during the transmission, then $\frac{R_{11}}{R_{21}}=\frac{R_{12}}{R_{22}}$, which is a special case of this general setting; this special case is important in practice as we expect that the channels between the transmitter and the receivers remain the same over the two rounds of transmission. We will only consider the two-user two-layer system in this work.

The notion of successive refinement of information in the single user setting was introduced by
Koshelev \cite{Koshelev:80} and by Equitz and Cover
\cite{EquitzCover:91} (see also \cite{Rimoldi:94}), and the problem is well researched. The main question is whether the requirement of encoding a source progressively necessitates a higher rate than encoding without the progressive requirement. A source is successively refinable if
encoding in multiple stages incurs no rate loss as compared with 
optimal rate-distortion encoding at the separate distortion levels; i.e., when the necessary encoding rate does not increase comparing to a single stage coding. The reassuring result by Equitz and Cover is that many familiar sources, such as the Gaussian source under the mean squared error (MSE) distortion measure and discrete sources under Hamming distortion measure, are in fact successively refinable. Lastras and Berger \cite{Lastras:01} further showed that when the source has a real alphabet and the distortion measure is MSE, even when the source is not successively refinable, the rate loss is bounded by a universal constant. 

In the multiuser setting we are interested in understanding whether the progressive coding requirement necessities any performance loss, and if so, whether the loss is bounded. In this work, we provide an achievable rate-distortion region for the problem with two users and two layers by embedding a (two-description) multiple description problem\footnote{The MSR problem is in fact a special case of the general problem of multiple descriptions with more than two descriptions; more precisely, there are four descriptions in the systems, but only four distortion requirements are considered, instead of one distortion constraint for each non-empty subset of the four descriptions. From here on, we shall use MD to stand for the conventional two description problem instead of the more general setting, unless specified otherwise explicitly. } inside it, and show that this region is tight for the Gaussian case. Furthermore, the loss of performance to a single layer coding can indeed be \textbf{unbounded}, which suggests unless there is a significant reason calling for a progressive coding, the loss of performance makes it a less attractive system design. 

The MSR problem includes the problem of multiple descriptions (MD) as a subsystem, and the techniques in the MD literature (notably \cite{Ozarow:80} and \cite{Gam:82}), are our main tools in this work. We show that the coding scheme given in \cite{Pradhanramchandran:02} based on random binning is sub-optimal, because multiple Gaussian side informations only at the decoders incur performance loss compared to the side information also available at the encoder; this is in contrast to the case of single side information at the decoder, where there is no essential loss \cite{Wynerziv:76}.
 The MSR problem is also related to the problem considered in \cite{Perron:06}, where each decoder has its individual side information, and the encoder has the complete set of side informations. 
The MSR problem further includes several variations of the MD problem, such as the MD problem with central refinement (MDCR) \cite{TianChen:05}, as well as the conditional MD problem. We will discuss  the MDCR problem in detail and reveal the implication of the general result from MSR. 

The distortion-rate (D-R) region and the rate-distortion (R-D) region given for the Gaussian MSR problem can be easily reduced to those for the MD problem. Though the Gaussian MD region has been known for more than 25 years, as pointed out in \cite{FengEffros:031} (discussed more extensively in \cite{FengEffros:07}), the expressions given in the literature are usually not complete (even incorrect if being used without caution) and we hope this confusion can be clarified by the present work\footnote{Feng and Effros clarified the Gaussian MD region in terms of R-D characterization in \cite{FengEffros:031}, but the interpretation of the degenerate region was not made explicit.}. 

It is worth pointing out that the formulation considered in this work is from the source coding point of view, which implies a coding system where source and channel coding are separated. However, it is well known that under the degraded broadcast channel, for which the considered coding approach is arguably the most suitable, a source-channel separation approach is not optimal (see, for example \cite{Gastpar:03}). 
It is nevertheless useful to consider the current formulation, since firstly joint source-channel coding (JSCC) schemes are often more complex, and secondly the performance using source-channel separation can be used to compare with that of any JSCC schemes to measure the possible performance loss. Moreover, in \cite{SteinbergMerhav:06} and \cite{TianDiggavi:072}, it was shown a source channel separation indeed holds in the scenario of successive refinement coding with side information at the decoder.  

The rest of the paper is organized as follows. In Section \ref{sec:problem} the problem is formally defined and some related background is given. An achievable region is given in Section \ref{sec:achievable}. In Section \ref{sec:Gaussian}, we prove that the given achievable region is tight for the Gaussian source, then analyze the performance loss comparing with single layer coding and discuss a special case with fixed channel configuration. Section \ref{sec:variation} discusses the MDCR problem as a special case of the problem being treated and Section \ref{sec:conclusion} concludes the paper.

\section{Problem Definition}
\label{sec:problem}

Let $\mathcal{X}$ be a finite set and let $\mathcal{X}^n$ be the set
of all $n$-vectors with components in $\mathcal{X}$.  Denote an
arbitrary member of $\mathcal{X}^n$ as $x^n=(x_1,x_2,...,x_n)$, or
alternatively as $\vec{x}$; $(x_i,x_2,...,x_j)$ will also be written as $x_{i,...,j}$.
 Upper case is used for random variables and vectors. A
discrete memoryless source (DMS) $(\mathcal{X},P_X)$ is an infinite
sequence $\{X_i\}_{i=1}^{\infty}$ of independent copies of a random variable
$X$ in $\mathcal{X}$ with a generic distribution $P_X$ with
$P_X(x^n)=\prod_{i=1}^nP_X(x_i)$. Let $\hat{\mathcal{X}}$ be a finite reconstruction alphabet, and 
for simplicity we assume that the decoders all use this reconstruction alphabet.
Let $\mathit{d}:\mathcal{X}\times\hat{\mathcal{X}}\rightarrow
\left[0,\infty\right)$ be a distortion measure. The single-letter
distortion of a vector is defined as
\begin{eqnarray}
\mathit{d}(\vec{x},\vec{\hat{x}})=\frac{1}{n}\sum_{i=1}^n
\mathit{d}(x_i,\hat{x}_i), \quad \forall \vec{x}\in \mathcal{X}^n,
\quad \vec{\hat{x}}\in \hat{\mathcal{X}}^n.
\end{eqnarray}

\begin{figure}[tb]
  \centering 
\includegraphics[scale=0.4]{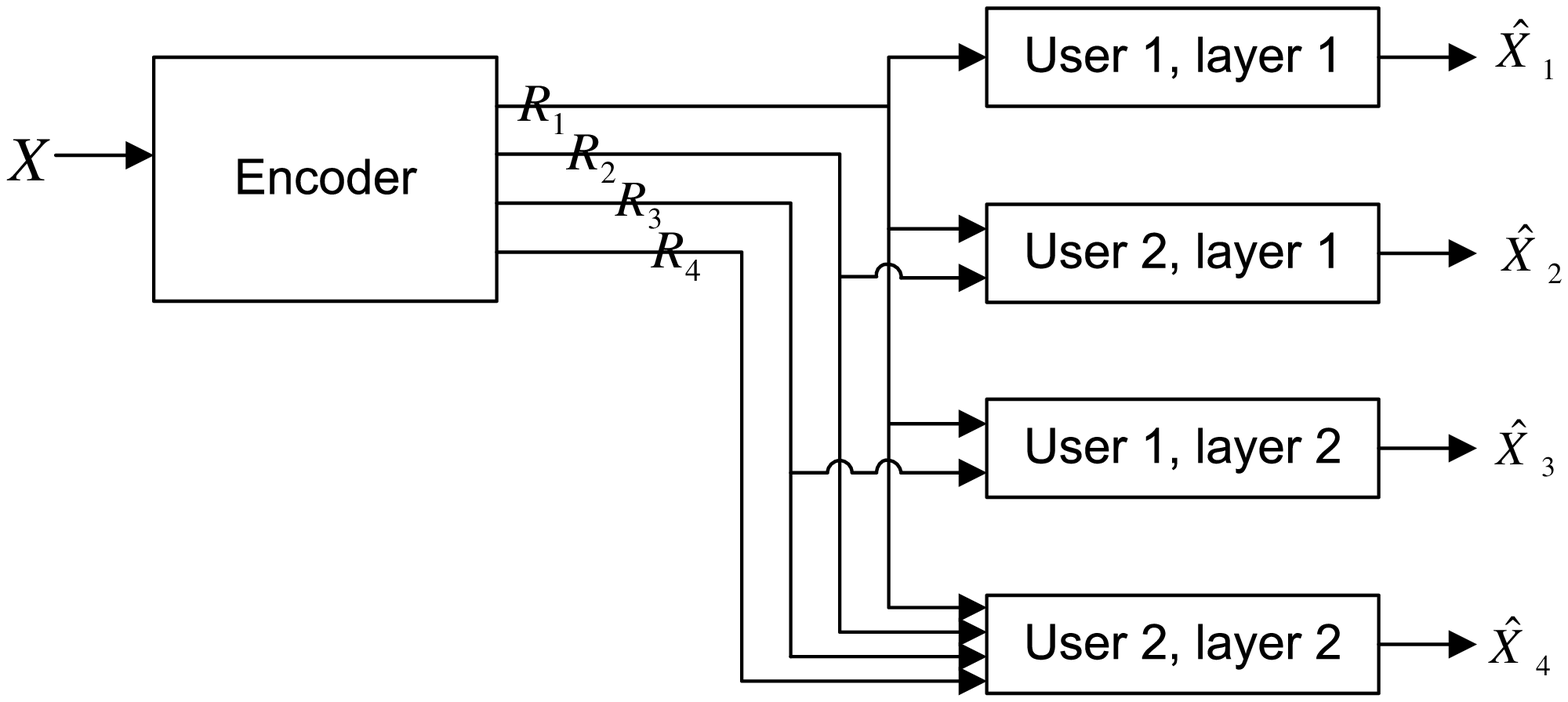}
\caption{\label{fig:eqlsystem} The equivalent system diagram to the two-user two-layer system depicted in Fig. \ref{fig:systemdiag}.}
\end{figure}

Instead of directly considering the system depicted in Fig. \ref{fig:systemdiag}, we consider the equivalent system given in Fig. \ref{fig:eqlsystem}. The reformulation is crucial, which makes the rather involved relations between descriptions more explicit. 
The double subscript in Fig. \ref{fig:systemdiag} is simplified to single subscript, whose correspondence is made clear in Table \ref{tab:labels}.

\begin{table}[tb]
\caption{The correspondence between system of Fig. \ref{fig:systemdiag} and that in Fig. \ref{fig:eqlsystem}.\label{tab:labels}}
\centering
\begin{tabular}{|c|c|c|c|c|}\hline
double subscript&(11)&(21)&(12)&(22)\\\hline 
single subscript&1   &2   & 3  &4 \\\hline 
\end{tabular}
\end{table}

\begin{definition}
An $(n,M_{1...4},D_{1...4})$ MSR code for source $X$ consists of 4 encoding
functions $\phi_i$ and 4 decoding functions $\psi_i$,
$i=1,2,3,4$:
\begin{eqnarray*}
\phi_i&:&\mathcal{X}^n\rightarrow I_{M_i},\,\,i=1,2,3,4
\end{eqnarray*}
where $I_k=\{1,2,...,k\}$ and 
\begin{eqnarray*}
\psi_1:I_{M_1} \rightarrow \hat{\mathcal{X}^n},&  \psi_2:I_{M_1}\times I_{M_2} \rightarrow \hat{\mathcal{X}^n},\\
\psi_3:I_{M_1}\times I_{M_3} \rightarrow \hat{\mathcal{X}^n},& 
\psi_4:I_{M_1}\times I_{M_2}\times I_{M_3}\times I_{M_4} \rightarrow \hat{\mathcal{X}^n},
\end{eqnarray*}
such that 
\begin{eqnarray*}
\Expt \mathit{d}(X^n,\psi_1(\phi_1(X^n)))&\leq& D_1,\\
\Expt \mathit{d}(X^n,\psi_2(\phi_1(X^n),\phi_2(X^n)))&\leq& D_2,\\
\Expt \mathit{d}(X^n,\psi_3(\phi_1(X^n),\phi_3(X^n)))&\leq& D_3,\\
\Expt \mathit{d}(X^n,\psi_4(\phi_1(X^n),\phi_2(X^n),\phi_3(X^n),\phi_4(X^n)))&\leq& D_4,
\end{eqnarray*}
where $\Expt$ is the expectation operation. 
\end{definition}

For the rest of the paper, we will often refer to the output of the encoding function $\phi_i$ as description $i$. All the logarithms and exponentials are base $e$. 

\begin{definition}
A rate distortion eight-tuple $(R_{1...4},D_{1...4})$ is said to be achievable, if for any
$\delta>0$ and sufficiently large $n$, there exist an
$(n,M_{1...4},D_1+\delta,D_2+\delta,D_3+\delta,D_4+\delta)$ MSR code, such that $R_i+\delta \geq \frac{1}{n}\log(M_i)$ for $i=1,2,3,4$.
\end{definition}

The MSR rate-distortion region, denoted by $\mathcal{Q}$, is the
set of all achievable eight-tuples. The problem of characterizing this region is difficult in general, because the problem at hand can be reduced to the well-known multiple description (MD) problem, which is a long standing open problem. In the MD problem, one sends two descriptions over two unreliable channels, either of which can break down; the question is to characterize the achievable rate-distortion quintuple consisting of the two description rates and the distortions of individual description, as well as that resulting from the two descriptions jointly. To reduce the MSR problem to the MD problem, we only need to set $R_1=R_4=0$ and $D_1=\infty$.

The literature on the MD problem is vast (see \cite{Goyal:011} for a review) and new results are emerging, but the problem remains open. Inner bound exists \cite{Gam:82} and it was shown that this bound is in fact quite good for source with real alphabet under MSE distortion measure \cite{Lastras:06}, but it is not tight in general \cite{ZhangBerger:87}. The Gaussian source with MSE distortion measure (and the recent extension to the vector Gaussian problem \cite{WangPramod:05}), for which this inner bound is tight, is the only case that the R-D region is completely characterized. Given these facts, our focus will not be on finding a complete solution for the general MSR problem. Instead, we will extend the coding scheme in \cite{Gam:82} to give an achievable region, and then focus on the quadratic Gaussian case, for which the achievable region is indeed tight.

We now briefly outline the coding scheme given by El Gamal and Cover in \cite{Gam:82} for the MD problem: given joint distribution $P_{XX_1X_2X_3}$, generate two length-$n$ codebooks using the marginals $P_{X_1}$ and $P_{X_2}$, respectively. It is well known that if approximately $\exp(nI(X;X_1))$ and $\exp(nI(X;X_2))$ codewords are generated for the two codebooks, respectively, then with high probability we can find codewords $X^n_1$, respectively $X^n_2$, jointly typical with any $X^n$ vector in the individual codebook. However, to guarantee the chosen codewords $X^n_1$, $X^n_2$ are also jointly typical together with $X^n$, the codebook sizes have to increase. The resulting increased rate is the expense paid to ``match" $X^n_1$ and $X^n_2$. Then for the matched $X^n_1$ and $X^n_2$ vector, a conditional codebook using $P_{X_3|X_1X_2}$ can be further added. The decoders then use the codewords $X^n_1$, $X^n_2$ and $X^n_3$ as the reconstructions.

\section{An Achievable Region}
\label{sec:achievable}

Several schemes were outlined in \cite{Pradhanramchandran:02} for the MSR problem which can achieve several specific operating points. One of them is to treat $\hat{X}_1$ and $\hat{X}_2$, which are the reconstructions in the first layer for user 1 and user 2, respectively, as side informations at the decoder and use the Wyner-Ziv binning approach \cite{Wynerziv:76} to generate description 3, and also use binning for description 4. The intuition behind this scheme is that the binning strategy has certain ``universal" property that whenever the bin is sufficiently small to decode with some side information, it can also be decoded with better quality side information (see \cite{Sgarro:77}). Since $\hat{X}_2$ is a better quality side information than $\hat{X}_1$, user 2 can also decode the description 3, which is meant for user 1; furthermore, user 2 can use it to improve its estimation.

Though the above observation is important, and perhaps provides the insight for the important result on symmetric $N$ description problem \cite{Pradhan:04}, it is not optimal for the current problem. Notice that since the receiver 2 has access to both description 1 and description 2, it can also reconstruct $\hat{X}_1$, in addition to $\hat{X}_2$ (which is its desired reconstruction). As such, a conditional codebook on $\hat{X}_1$ is more suitable, since it is available at both the encoder and the decoder. 

It should now be clear that the MD coding method can be used in MSR, if we treat $\hat{X}_1$ as the common side information available at both the encoder and the decoders when encoding for $\hat{X}_2$, $\hat{X}_3$ and  $\hat{X}_4$. To insure there are sufficient codewords in the $\hat{X}_1$ codebook such that source vectors are covered with high probability, a rate of $R_1=I(X;\hat{X}_1)+\delta_1$ can be chosen ($\delta_i,i=1,2,3$ are small positive quantities). 
Conditioned on $\hat{X}_1$, two codebooks of size $\exp(nR_2)$ and $\exp(nR_3)$, respectively, are generated using $P_{\hat{X}_2|\hat{X}_1}$ and $P_{\hat{X}_3|\hat{X}_1}$; as discussed in the last section, in order to find $\hat{X}_2^n$ and $\hat{X}_3^n$ (conditioned on $\hat{X}_1$) jointly typical with $X^n$ (i.e., matched) in these two codebooks with high probability, the codebook sizes should be chosen accordingly. More precisely, we can choose
\begin{eqnarray} 
&&R_2+R_3=I(X;\hat{X}_2\hat{X}_3|\hat{X}_1)+I(\hat{X}_2;\hat{X}_3|\hat{X}_1)+\delta_2,\\
&&R_2>I(X;\hat{X}_2|\hat{X}_1),\quad R_3>I(X;\hat{X}_3|\hat{X}_1).
\end{eqnarray} 
Given the codeword $\hat{X}_1^n$ and the matched codewords $\hat{X}_2^n$ and $\hat{X}_3^n$, in the last coding stage a codeword in the codebook of size $\exp(nR_4)$ generated by $P_{\hat{X}_4|\hat{X}_1\hat{X}_2\hat{X}_3}$ is chosen which is jointly typical with the source vector $X^n$ and the previously chosen codewords $\hat{X}_1^n,\hat{X}_2^n$ and $\hat{X}_3^n$; for such a purpose we can choose $R_4=I(X;\hat{X}_4|\hat{X}_1\hat{X}_2\hat{X}_3)+\delta_3$. Now an achievable region is readily available using standard techniques, though it is not clear if it is optimal. 

Here we would like to bring attention to a quite subtle and often-overlooked fact: even when $R_1=0$, the reduced problem is still not the same as the MD problem. Notice there are three rates $(R_2,R_3,R_4)$ to characterize here, instead of the two rates in the MD problem; it is nevertheless a special case of the general three description problem. This problem, which we refer to as the multiple descriptions with central refinement (MDCR) problem \cite{TianChen:05}, will be treated in more depth later. Though not the same, the MDCR system is not unfamiliar: the coding scheme in \cite{Gam:82} in fact uses such a structure.

Given the discussion above, we next state an achievable region without detailed proof for the sake of brevity\footnote{See also \cite{DiggaviVinay:04} for a similar result for multiple descriptions when both encoder and decoders have access to common side information.}. Define the region $\mathcal{Q}_{ach}$ to be the set of all rate distortion eight-tuples $(R_{1...4},D_{1...4})$ for which there exist four random variables $\hat{X}_1,\hat{X}_2,\hat{X}_3,\hat{X}_4$ in finite alphabet $\hat{\mathcal{X}}$ such that 
\begin{eqnarray}
\Expt\mathit{d}(X,\hat{X}_i)\leq D_i,\,\,i=1,2,3,4,
\end{eqnarray}
and the non-negative rate vector satisfies:
\begin{eqnarray}
R_1&\geq& I(X;\hat{X}_1)\\ 
\sum_{i=1,2}R_i&\geq& I(X;\hat{X}_1\hat{X}_2),\\
\sum_{i=1,3}R_i&\geq& I(X;\hat{X}_1\hat{X}_3),\\
\sum_{i=1,2,3}R_i&\geq&  I(X;\hat{X}_1\hat{X}_2\hat{X}_3)+I(\hat{X}_2;\hat{X}_3|\hat{X}_1),\\
\sum_{i=1,2,3,4}R_i&\geq& I(X;\hat{X}_1\hat{X}_2\hat{X}_3\hat{X}_4)+I(\hat{X}_2;\hat{X}_3|\hat{X}_1).
\end{eqnarray} 

The following theorem provides an achievable region. 

\begin{theorem}
\label{theorem:achievable}
\begin{eqnarray*}
\mathcal{Q}_{ach}\subseteq \mathcal{Q}.
\end{eqnarray*}
\end{theorem}

If $R_1=R_4=0$, $\mathcal{Q}_{ach}$ degenerates to the achievable region given by El Gamal and Cover in \cite{Gam:82}. The region is characterized by a set of sum-rate bounds, instead of the individual rate for $(R_1,R_2,R_3,R_4)$. It is not immediately clear that the aforementioned coding scheme (with individual rates) can achieve the complete region characterized by the sum-rates in Theorem 1. However, a moment of thought reveals that the structure of this system implies that for any code with rates $(r_1,r_2,r_3,r_4)$, we can freely move the rates $r_2$ (and $r_3$) into $r_1$, and rate $r_4$ into $r_1,r_2,r_3$ to construct new codes; see \cite{Effros:99} (also \cite{Rimoldi:94}) for a thorough explanation regarding a similar property in the successive refinement problem. Thus indeed the region given in Theorem 1 is achievable.

It is not clear whether the region $\mathcal{Q}_{ach}$ is convex. Interestingly, a generalization of this region, denoted as $\mathcal{Q}'_{ach}$, is indeed convex, and we now provide this generalized region. Let the region $\mathcal{Q}'_{ach}$ be the set of all rate distortion eight-tuples $(R_{1...4},D_{1...4})$ for which there exist four random variables $U_1,U_2,U_3,U_4$ in finite alphabets $\mathcal{U}_1,\mathcal{U}_2,\mathcal{U}_3,\mathcal{U}_4$ such that there exists deterministic functions 
\begin{eqnarray}
g_1:\mathcal{U}_1\rightarrow \hat{\mathcal{X}},&\quad& g_2:\mathcal{U}_1\times\mathcal{U}_2\rightarrow \hat{\mathcal{X}}\\
g_3:\mathcal{U}_1\times\mathcal{U}_3\rightarrow \hat{\mathcal{X}},&\quad& g_4:\mathcal{U}_1\times\mathcal{U}_2\times\mathcal{U}_3\times\mathcal{U}_4\rightarrow \hat{\mathcal{X}}
\end{eqnarray}
satisfying
\begin{eqnarray}
\Expt\mathit{d}(X,g_1(U_1))\leq D_1,& \Expt\mathit{d}(X,g_2(U_1,U_2))\leq D_2,\\
\Expt\mathit{d}(X,g_3(U_1,U_3))\leq D_3,& \Expt\mathit{d}(X,g_4(U_1,U_2,U_3,U_4))\leq D_4
\end{eqnarray}
and the non-negative rate vector satisfies:
\begin{eqnarray}
\label{eqn:rates}
R_1&\geq& I(X;U_1), \\ 
\sum_{i=1,2}R_i&\geq& I(X;U_1U_2),\\
\sum_{i=1,3}R_i&\geq& I(X;U_1U_3),\\
\sum_{i=1,2,3}R_i&\geq&  I(X;U_1U_2U_3)+I(U_2;U_3|U_1),\label{eqn:ratesum}\\
\sum_{i=1,2,3,4}R_i&\geq& I(X;U_1U_2U_3U_4)+I(U_2;U_3|U_1).
\end{eqnarray} 

\textit{\textbf{Theorem}} 1$^\prime$:
\begin{eqnarray*}
\mathcal{Q}'_{ach}\subseteq \mathcal{Q}.
\end{eqnarray*}

The proof of this theorem follows the exact same line as that of Theorem \ref{theorem:achievable}. Moreover, it can be shown straightforwardly that $\mathcal{Q}_{ach}\subseteq\mathcal{Q}'_{ach}$, but it is not clear whether the inclusion in the other direction is also true. Though the region $\mathcal{Q}'_{ach}$ is more general, it is also more complex due to the involvement of several decoding  functions. In fact, as suggested in \cite{ZhangBerger:87}, the achievable region given by El Gamal and Cover \cite{Gam:82} originally had a form with several decoding functions, which was later largely abandoned in favor of the region defined without such functions\footnote{Different from the MSR problem, in the MD problem, the region defined without the decoding function is in fact more general; see \cite{ZhangBerger:87}.}. The region $\mathcal{Q}'_{ach}$ is more suitable when we consider the Gaussian source, and we thus include both forms here for completeness.

It is not difficult to show that $\mathcal{Q}'_{ach}$ is convex. Let $P(U^0_1,U^0_2,U^0_3,U^0_4|X)$ and $P(U^1_1,U^1_2,U^1_3,U^1_4|X)$ be two conditional distributions which provide rate vectors in  $\mathcal{Q}'_{ach}$. Let $Q$ be a Bernoulli random variable with $\mbox{Pr}(Q=0)=\lambda$ and $\mbox{Pr}(Q=1)=1-\lambda$, which is independent of everything, then it is easily seen
\begin{eqnarray}
&&\lambda I(X;U^0_1)+(1-\lambda)I(X;U^1_1)\nonumber\\
&=&\lambda I(X;U^0_1|Q=0)+(1-\lambda)I(X;U^1_1|Q=1)\\
&=&I(X;U^Q_1,Q).
\end{eqnarray} 
Moreover 
\begin{align}
\lambda I(U^0_2;U^0_3|U^0_1)+(1-\lambda)I(U^1_2;U^1_3|U^1_1)=I(U^Q_2;U^Q_3|U^Q_1,Q).
\end{align}
Similar relations can be derived for the other mutual information quantities. Now define $U_i=(U^Q_i,Q)$, and the decoding functions can be defined accordingly; it follows that this convex combination of the rate vectors is indeed in $\mathcal{Q}'_{ach}$.

Though we have considered discrete sources so far, the results can be generalized to Gaussian sources using the techniques in
\cite{Gallagerbook}\cite{Wyner:78}. In the next section, we prove a converse for the Gaussian source under the MSE distortion measure, and show that the region given in Theorem \ref{theorem:achievable} is tight for the Gaussian source. It is worth clarifying that the converse result for the Gaussian MSR problem is not implied by that of the Gaussian source with common side information at both the encoder and the decoders \cite{DiggaviVinay:04}, because the optimal first codebook in MSR is not necessarily a codebook generated with any single letter marginal distribution $P_{\hat{X}_1}$, while the common side information is always an i.i.d. random variable in the setting of \cite{DiggaviVinay:04}; moreover, the MSR problem is further complicated by the included MDCR sub-system. 

\section{The Gaussian Source}
\label{sec:Gaussian}
In this section, we shall focus on the Gaussian source with MSE distortion measure, and establish the distortion-rate as well as the rate-distortion region. 

\subsection{The distortion-rate region for the Gaussian source}
\label{subsec:converse}
Theorem 2 below gives the distortion-rate (D-R) region for the Gaussian source, and Theorem 2$^\prime$ gives the rate-distortion (R-D) region. Similar to the MD problem, the D-R region is simpler than the R-D region due to less number of degenerate regions (see \cite{FengEffros:031} and \cite{ChenTian:05}). It will be illustrated that the R-D region can be established from the D-R region. We shall first present the theorems, and then follow the approach by Ozarow \cite{Ozarow:80} to establish the D-R region. 

\begin{theorem}
\label{theorem:Gaussian}
For the Gaussian source $X\sim \mathcal{N}(0,\sigma_x^2)$ under MSE distortion measure, the achievable distortion-rate region for rates $(R_1,R_2,R_3,R_4)$ is given by 
\begin{eqnarray}
\label{eqn:DR}
d_1&\geq& \sigma_x^2\exp[-2R_1],\\
d_2&\geq& \sigma_x^2\exp[-2(R_1+R_2)],\\
d_3&\geq& \sigma_x^2\exp[-2(R_1+R_3)],\label{eqn:DR3}\\
d_4&\geq& \frac{\sigma_x^2\exp[-2(R_1+R_2+R_3+R_4)]}{1-(|\sqrt{\Pi}-\sqrt{\Delta}|^+)^2}\label{eqn:DR4}
\end{eqnarray}
where $|x|^+=\max(x,0)$ and 
\begin{align}
d_1^*\stackrel{\Delta}{=}\sigma_x^2\exp[-2R_1],\quad\hat{d_2}\stackrel{\Delta}{=}\min(d_2,d_1^*), \quad \hat{d_3}\stackrel{\Delta}{=}\min(d_3,d_1^*)\nonumber\\
\quad \Pi\stackrel{\Delta}{=}(1-\frac{\hat{d}_2}{d_1^*})(1-\frac{\hat{d}_3}{d_1^*}),\quad\Delta\stackrel{\Delta}{=}\frac{\hat{d}_2\hat{d}_3}{d_1^{*2}}-\exp[-2(R_2+R_3)].
\end{align}
\end{theorem}

Define the function $R(D)=\frac{1}{2}\log\frac{1}{D}$. The following theorem describes the R-D region for the Gaussian source.

\textit{\textbf{Theorem}} 2$^\prime$: For the Gaussian source $X\sim \mathcal{N}(0,\sigma_x^2)$ under MSE distortion measure, the achievable rate-distortion region for distortions $(d_1,d_2,d_3,d_4)$ is given as follows. 
\begin{eqnarray}
R_1\geq R_1^*\stackrel{\Delta}{=}R(\frac{\min(d_1,\sigma_x^2)}{\sigma_x^2}). 
\label{eqn:RDR1}
\end{eqnarray}
For any rates $R_1\geq R_1^*$ and $R_4\geq 0$, define $\hat{d_4}\stackrel{\Delta}{=}d_4\exp(2R_4)$. The achievable rates $(R_2,R_3)$ are given as: 
\begin{eqnarray}
\label{eqn:RDR2R3}
R_2\geq R(\hat{d_2}/d_1^*),\quad R_3\geq R(\hat{d_3}/d_1^*),
\end{eqnarray}
\begin{align}
R_2+R_3\geq \left\{\begin{array}{lr}
              \textstyle{R(\hat{d_4}/d_1^*)} & 0<\hat{d_4}<\hat{d_2}+\hat{d_3}-d_1^*;\\
              \textstyle{0} &\hat{d_4}>(\hat{d_2}^{-1}+\hat{d_3}^{-1}-(d_1^*)^{-1})^{-1} ;\\              
              \textstyle{R(\hat{d_4}/d_1^*)+L} & \mbox{otherwise},\\
              \end{array}\right.   \label{eqn:RD}
\end{align}
where
\begin{eqnarray}
L=\frac{1}{2}\log\frac{(d_1^*-\hat{d_4})^2}{(d_1^*-\hat{d_4})^2+(d_1^*\sqrt{\Pi}+\sqrt{\hat{d_2}-\hat{d_4}}\sqrt{\hat{d_3}-\hat{d_4}})^2}.
\end{eqnarray}

There is one degenerate case in the D-R region (when $\Pi<\Delta$), and there are two degenerate cases in the R-D region (the first two cases in (\ref{eqn:RD})). They are degenerate in the sense that any  eight-tuple in those regions is worse than or equal to (in each component) an eight-tuple on the boundary of the non-degenerate region. This interpretation is made more explicit at the end of the forward proof for Theorem 2.    

\begin{figure*}[htb]
  \centering 
\includegraphics[scale=0.5]{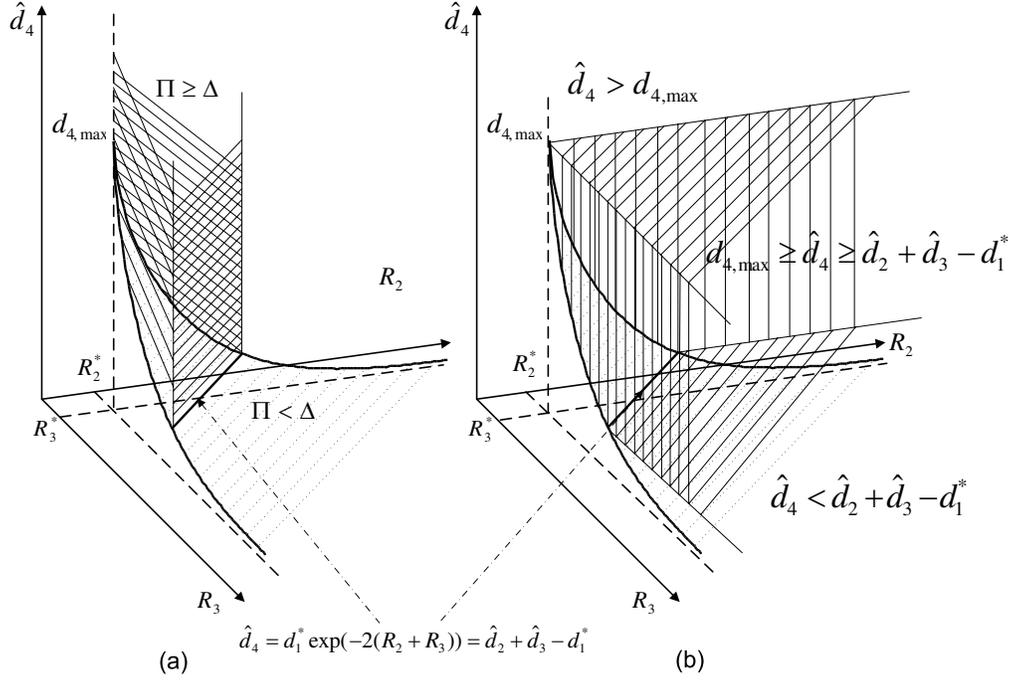}
\caption{The equivalence of the R-D and D-R characterization of $(R_2,R_3,\hat{d}_4)$ (a) D-R characterization (b) R-D characterization.\label{fig:equivalence} $R_2^*=R(\hat{d}_2/d_1^*)$, $R_3^*=R(\hat{d}_3/d_1^*)$, and $d_{4,\max}=(\hat{d}_2^{-1}+\hat{d}_3^{-1}-(d_1^*)^{-1})^{-1}$}
\end{figure*}

The region given in Theorem \ref{theorem:Gaussian} reduces to the Gaussian MD region, when $R_1=R_4=0$. The form of this achievable region is not surprising, given the aforementioned achievable scheme and the Guassian MD region in \cite{Ozarow:80} (with the additional degenerate case made explicit here). However, the converse is not yet clear due to the involvement of the coding functions $\phi_1$ and $\phi_4$. More precisely, the following two questions regarding $\phi_1$ and $\phi_4$, respectively, can be asked:
\begin{enumerate}
\item Is a Gaussian codebook optimal for encoder $\phi_1$?
\item With the additional information provided by $\phi_4$, should the codebooks generated for $\phi_2$ and $\phi_3$ still have the same structure as the MD codebooks? In other words, should there still be matched codewords for (almost) every typical source sequence in the codebooks for $\phi_2$ and $\phi_3$ (or the matching can be moved into $\phi_4$)?
\end{enumerate}
In the proof, we will show that the answers to both the questions are positive. The main difference from the well-known proof by Ozarow \cite{Ozarow:80} for the Gaussian MD problem is the additional coding stages $\phi_1$ and $\phi_4$, which makes the converse more involved, and the entropy power inequality has to be applied twice in the proof.

Before proceeding to the proof of Theorem \ref{theorem:Gaussian}, we illustrate the equivalence of the D-R and the R-D characterizations. The inequalities in (\ref{eqn:DR})-(\ref{eqn:DR3}) are clearly equivalent to (\ref{eqn:RDR1}) and the two inequalities in (\ref{eqn:RDR2R3}). Thus we only need to establish that with them, the region of triples $(R_2,R_3,\hat{d}_4)$ characterized by (\ref{eqn:DR4}) and that characterized by (\ref{eqn:RD}) are equivalent for any valid and fixed $(R_1,R_4,d_1,d_2,d_3)$. 
This is illustrated geometrically in Fig. \ref{fig:equivalence}. It is seen that the same region above the surface can be described either in two regimes as in the D-R characterization, or in three regimes as in the R-D characterization. The functions given in Theorem 2 and Theorem 2$^\prime$ specifying the regions can be shown to be equivalent with some amount of algebra (see \cite{FengEffros:031} and \cite{FengEffros:07} for a brief discussion on this algebraic computation in the MD problem). Note that the same argument is true for the MD problem, simply by taking $R_1=R_4=0$.

\begin{proof}[Theorem \ref{theorem:Gaussian}]

\noindent\textbf{Converse:} The following bounds are straightforward by conventional rate-distortion theory: 
\begin{eqnarray}
\label{eqn:easybound}
d_1&\geq& \sigma_x^2\exp(-2R_1),\nonumber\\
d_2&\geq& \sigma_x^2\exp[-2(R_1+R_2)],\nonumber\\
d_3&\geq& \sigma_x^2\exp[-2(R_1+R_3)].
\end{eqnarray}

We also have 
\begin{eqnarray*}
I(X^n;\hat{X}^n_4)&\stackrel{(a)}{\leq}& I(X^n;\phi_1\phi_2\phi_3\phi_4)\leq H(\phi_1\phi_2\phi_3\phi_4)\\
&=&H(\phi_1)+H(\phi_2|\phi_1)+H(\phi_3|\phi_1)\\
&&\qquad\qquad-I(\phi_2;\phi_3|\phi_1)+H(\phi_4|\phi_1\phi_2\phi_3)\\
&\stackrel{(b)}{\leq}& n(R_1+R_2+R_3+R_4)-I(\phi_2;\phi_3|\phi_1),
\end{eqnarray*}
where (a) is because $\hat{X}^n_4$ is determined by the encoding functions $(\phi_1,\phi_2,\phi_3,\phi_4)$, and (b) is because conditioning reduces entropy as $H(\phi_3|\phi_1)\leq H(\phi_3)$, $H(\phi_2|\phi_1)\leq H(\phi_2)$ and $H(\phi_4|\phi_1\phi_2\phi_3)\leq H(\phi_4)$, and because of the cardinalities of the encoding functions.
By converse to the source coding theorem
\begin{eqnarray}
d_4&\geq& D(\frac{1}{n}I(X^n;\hat{X}^n_4))\nonumber\\
&\geq& \sigma_x^2\exp[-2(R_1+R_2+R_3+R_4)]\exp[\frac{2}{n}I(\phi_2;\phi_3|\phi_1)].\nonumber\\
&&\label{eqn:d4}
\end{eqnarray}
Because $\hat{X}^n_2$ and $\hat{X}^n_3$  are functions of $\phi_1,\phi_2$ and $\phi_1,\phi_3$, respectively, it is seen that 
\begin{eqnarray}
I(\phi_2;\phi_3|\phi_1)\geq I(\hat{X}^n_2;\hat{X}^n_3|\phi_1).\label{eqn:If1f2}
\end{eqnarray}

As in the proof by Ozarow \cite{Ozarow:80} for the MD problem, let $Y=X+N$, where $N$ is zero mean Gaussian with variance $\epsilon$ and independent of $X$.
Because of the following identity
\begin{eqnarray*}
&&I(\hat{X}^n_2;\hat{X}^n_3Y^n|\phi_1)\\
&=&I(\hat{X}^n_2;\hat{X}^n_3|Y^n\phi_1)+I(\hat{X}^n_2;Y^n|\phi_1)\\
&=&I(\hat{X}^n_2;\hat{X}^n_3|\phi_1)+I(\hat{X}^n_2;Y^n|\hat{X}^n_3\phi_1)\\
&=&I(\hat{X}^n_2;\hat{X}^n_3|\phi_1)+I(\hat{X}^n_2\hat{X}^n_3;Y^n|\phi_1)-I(\hat{X}^n_3;Y^n|\phi_1),
\end{eqnarray*}
we have 
\begin{eqnarray}
&&I(\hat{X}^n_2;\hat{X}^n_3|\phi_1)\nonumber\\
&\geq& I(\hat{X}^n_2;Y^n|\phi_1)+I(\hat{X}^n_3;Y^n|\phi_1)-I(\hat{X}^n_2\hat{X}^n_3;Y^n|\phi_1)\nonumber\\
&=&h(Y^n|\phi_1)-h(Y^n|\hat{X}^n_2\phi_1)+h(Y^n|\phi_1)\nonumber\\
&&\qquad-h(Y^n|\hat{X}^n_3\phi_1)-h(Y^n|\phi_1)+h(Y^n|\hat{X}^n_2\hat{X}^n_3\phi_1)\nonumber\\
&=&h(Y^n|\phi_1)-h(Y^n|\hat{X}^n_2\phi_1)-h(Y^n|\hat{X}^n_3\phi_1)\nonumber\\
&&\qquad\qquad\qquad\qquad\qquad\qquad\qquad\,+h(Y^n|\hat{X}^n_2\hat{X}^n_3\phi_1)\nonumber\\
&=&I(Y^n;\hat{X}^n_2\phi_1)+I(Y^n;\hat{X}^n_3\phi_1)-2h(Y^n)\nonumber\\
&&\qquad\qquad\qquad\qquad\quad+h(Y^n|\phi_1)+h(Y^n|\hat{X}^n_2\hat{X}^n_3\phi_1)\nonumber\\
&\geq&I(Y^n;\hat{X}^n_2)+I(Y^n;\hat{X}^n_3)-2h(Y^n)\nonumber\\
&&\qquad\qquad\qquad\qquad+h(Y^n|\phi_1)+h(Y^n|\hat{X}^n_2\hat{X}^n_3\phi_1).\label{eqn:IX2X3}
\end{eqnarray}

Taking $Y$ as a Gaussian source, the distortion between $Y^n$ and $\hat{X}^n_2$ is upper bounded by $n(d_2+\epsilon)$, by the converse to source coding theorem
\begin{eqnarray}
\label{eqn:a1}
I(Y^n;\hat{X}^n_2)\geq nR_Y(d_2+\epsilon)=\frac{n}{2}\log\frac{\sigma_x^2+\epsilon}{d_2+\epsilon},
\end{eqnarray}
where $R_Y(D)$ is the rate distortion function for source $Y$. 
Similarly
\begin{eqnarray}
\label{eqn:a2}
I(Y^n;\hat{X}^n_3)\geq \frac{n}{2}\log\frac{\sigma_x^2+\epsilon}{d_3+\epsilon}.
\end{eqnarray}
The following steps in our converse proof are different from Ozarow's, and it is worth noting the complication introduced by the coding function $\phi_1$. 
We apply the conditional entropy power inequality  \cite{CoverThomas} on the term $h(Y^n|\phi_1)$, which gives
\begin{eqnarray}
h(Y^n|\phi_1)\geq \frac{n}{2}\log[\exp(\frac{2}{n}h(X^n|\phi_1))+2\pi e \epsilon].
\end{eqnarray}
However notice that 
\begin{eqnarray*}
h(X^n|\phi_1)=h(X^n)-I(X^n;\phi_1)\geq \frac{n}{2}\log(2\pi e\sigma_x^2)-nR_1,
\end{eqnarray*}
which gives
\begin{eqnarray}
h(Y^n|\phi_1)&\geq& \frac{n}{2}\log[\exp(\frac{2}{n}[\frac{n}{2}\log(2\pi e\sigma_x^2)-nR_1])+2\pi e \epsilon]\nonumber\\
&=&\frac{n}{2}\log[2\pi e(\sigma_x^2\exp(-2R_1)+\epsilon)].\label{eqn:HYn}
\end{eqnarray}
Applying the entropy power inequality again on the term $h(Y^n|\hat{X}^n_2\hat{X}^n_3\phi_1)$, we have
\begin{eqnarray*}
h(Y^n|\hat{X}^n_2\hat{X}^n_3\phi_1)\geq \frac{n}{2}\log [\exp(\frac{2}{n}h(X^n|\hat{X}^n_2\hat{X}^n_3\phi_1))+2\pi e \epsilon].
\end{eqnarray*}
It follows that 
\begin{eqnarray*}
&&h(X^n|\hat{X}^n_2\hat{X}^n_3\phi_1)\nonumber\\
&=&h(X^n|\phi_1)-I(X^n;\hat{X}^n_2\hat{X}^n_3|\phi_1)\\
&\stackrel{(a)}{=}&h(X^n|\phi_1)-H(\hat{X}^n_2\hat{X}^n_3|\phi_1)\\
&=&h(X^n)-I(X^n;\phi_1)-H(\hat{X}^n_2|\phi_1)\nonumber\\
&&\qquad\qquad\qquad\qquad-H(\hat{X}^n_3|\phi_1)+I(\hat{X}^n_2;\hat{X}^n_3|\phi_1)\\
&\stackrel{(b)}{\geq}&h(X^n)-I(X^n;\phi_1)-H(\phi_2|\phi_1)\nonumber\\
&&\qquad\qquad\qquad\qquad-H(\phi_3|\phi_1)+I(\hat{X}^n_2;\hat{X}^n_3|\phi_1)\\
&\geq&\frac{n}{2}\log (2\pi e \sigma^2_x)-n(R_1+R_2+R_3)+I(\hat{X}^n_2;\hat{X}^n_3|\phi_1),
\end{eqnarray*}
where (a) follows from the fact that $\hat{X}^n_2$ and $\hat{X}^n_3$ are functions of $X^n$, and (b) from the fact $\hat{X}^n_2$ and $\hat{X}^n_3$ are functions of $(\phi_1,\phi_2)$ and $(\phi_1,\phi_3)$, respectively. 
This leads to 
\begin{align}
&h(Y^n|\hat{X}^n_2\hat{X}^n_3\phi_1)\nonumber\\
&\geq \frac{n}{2}\log \left\{2\pi e \epsilon\vphantom{+\exp\left(\frac{2}{n}[\frac{n}{2}\log (2\pi e  \sigma^2_x)-n\sum_{i=1}^3R_i+I(\hat{X}^n_2;\hat{X}^n_3|\phi_1)] \right)}\right.\nonumber\\
&\qquad\,\,\,\left.+\exp\left(\frac{2}{n}[\frac{n}{2}\log (2\pi e  \sigma^2_x)-n\sum_{i=1}^3R_i+I(\hat{X}^n_2;\hat{X}^n_3|\phi_1)] \right)\right\}\nonumber\\
&=\frac{n}{2}\log\left\{2\pi e \left(\sigma^2_x\exp[-2\sum_{i=1}^3R_i]\exp[\frac{2}{n}I(\hat{X}^n_2;\hat{X}^n_3|\phi_1)]+\epsilon\right)\right\}.\label{eqn:HYnX2X3}
\end{align}
Define 
\begin{eqnarray}
t\stackrel{\Delta}{=}\exp[\frac{2}{n}I(\hat{X}^n_2;\hat{X}^n_3|\phi_1)],
\label{eqn:definet}
\end{eqnarray}
and summarize all the bounds in (\ref{eqn:IX2X3}), (\ref{eqn:a1}), (\ref{eqn:a2}), (\ref{eqn:HYn}) and (\ref{eqn:HYnX2X3}), and we thus have
\begin{eqnarray}
t\geq \frac{d_1^*+\epsilon}{(d_2+\epsilon)(d_3+\epsilon)}[td_1^*\exp[-2(R_2+R_3)]+\epsilon].
\end{eqnarray}
Isolating $t$ we have
\begin{eqnarray}
t\geq \frac{\epsilon(d_1^*+\epsilon)}{(d_2+\epsilon)(d_3+\epsilon)-(d_1^*+\epsilon)d_1^*\exp[-2(R_2+R_3)]},
\end{eqnarray}
notice that because $d_2\geq d_1^*\exp(-2R_2)$ and $d_3\geq d_1^*\exp(-2R_3)$ from (\ref{eqn:easybound}), the denominator is always positive, as long as $\epsilon$ is positive. 

To get the tightest bound, we maximize the lower bound on $t$ over $\epsilon$. When $d_2\leq d^*_1$ and $d_3\leq d^*_1$, define 
\begin{eqnarray}
\Pi^*=(1-\frac{d_2}{d_1^*})(1-\frac{d_3}{d_1^*}),\quad
\Delta^*=\frac{d_2d_3}{d_1^{*2}}-\exp[-2(R_2+R_3)].
\end{eqnarray}
Then we choose the following value of $\epsilon$
\begin{eqnarray}
\epsilon&=&\left\{\begin{array}{ll}
              \ \frac{d_1^*\sqrt{\Delta^*}}{\sqrt{\Pi^*}-\sqrt{\Delta^*}}& \Pi^*\geq \Delta^*;\\
              \ \infty &\text{otherwise}.\\
              \end{array}\right.  
\end{eqnarray}
After some algebraic calculation, we have for the case $\Pi^*\geq \Delta^*$, 
\begin{eqnarray}
t\geq \frac{1}{1-(\sqrt{\Pi^*}-\sqrt{\Delta^*})^2},\label{eqn:t}
\end{eqnarray} 
and subsequently using (\ref{eqn:d4}), (\ref{eqn:If1f2}), (\ref{eqn:definet}) and (\ref{eqn:t})
\begin{eqnarray}
d_4\geq \frac{\sigma_x^2\exp[-2(R_1+R_2+R_3+R_4)]}{1-(\sqrt{\Pi^*}-\sqrt{\Delta^*})^2}.
\end{eqnarray}
For the case $\Pi^*< \Delta^*$, we have $t\geq 1$, which gives the trivial bound of 
\begin{eqnarray}
d_4\geq \sigma_x^2\exp[-2(R_1+R_2+R_3+R_4)].
\label{eqn:trivialbound}
\end{eqnarray}
This is not yet the bound given in Theorem \ref{theorem:Gaussian} since the definitions of $\Pi$ and $\Delta$ are not the same as those of $\Pi^*$ and $\Delta^*$, which are only defined when $d_2\leq d^*_1$ and $d_3\leq d^*_1$. To close this gap, note that if $d_2\geq d^*_1$ (or $d_3\geq d^*_1$), we may trivially write the lower bound (\ref{eqn:trivialbound}), which coincides with (\ref{eqn:DR4}) for this case, due to the fact $\Pi=0$ by the definition of $\hat{d}_2$ (or $\hat{d}_3$). Thus the lower bound in Theorem \ref{theorem:Gaussian} is established.

\vspace{0.5cm}
\noindent\textbf{Forward:} 
Now we shall use the general achievable region given in Theorem $1'$ to derive an inner bound for the Gaussian source, and show it coincides with the outer bound.
 
Construct the following random variables
\begin{align}
&U_1=X+N_1, X'=X-\Expt(X|U_1), U_2=X'+N_2,\\ &U_3=X'+N_3,\quad U_4=X'+N_4,
\end{align}
where $N_1,N_2,N_3,N_4$ are zero mean jointly Gaussian, independent of $X$, and having the covariance matrix 
\begin{eqnarray}
\begin{pmatrix}
\sigma_1^2 &0 & 0 & 0\\
0     &\sigma_2^2 & \rho\sigma_2\sigma_3 &0\\
0     &\rho\sigma_2\sigma_3& \sigma_3^3  &0\\
0     &0          &0                     &\sigma_4^2       
\end{pmatrix}
\end{eqnarray}
It can be seen that $X'$ is essentially the innovation of $X$ given $U_1$. 
The decoding functions are 
\begin{eqnarray}
\hat{X}_1&=&f_1(U_1)=\Expt(X|U_1)=\frac{\sigma_x^2}{\sigma_x^2+\sigma_1^2}U_1,\\
\hat{X}_2&=&f_2(U_1,U_2)=\hat{X}_1+\Expt(X'|U_2)\\
\hat{X}_3&=&f_3(U_1,U_3)=\hat{X}_1+\Expt(X'|U_3),\\
\hat{X}_4&=&f_4(U_1,U_2,U_3,U_4)=\hat{X}_1+\Expt(X'|U_2U_3U_4).
\end{eqnarray}

We have that the following rate $R_1$ is achievable
\begin{eqnarray}
R_1\geq I(X;U_1)=\frac{1}{2}\log\frac{\sigma_x^2+\sigma_1^2}{\sigma_1^2}.
\end{eqnarray}
Choose $\sigma_1$ such that the above inequality holds with equality. 
Then we have 
\begin{eqnarray}
d_1=\Expt(X'^2)=\frac{\sigma_1^2\sigma_x^2}{\sigma_1^2+\sigma_x^2}=\sigma_x^2\exp(-2R_1)=d^*_1.
\end{eqnarray}
We have also 
\begin{eqnarray}
d_2=\Expt[X-\hat{X}_2]^2=\Expt[X'-\Expt(X'|U_2)]^2=\frac{d_1\sigma_2^2}{d_1+\sigma_2^2},\\
d_3=\Expt[X-\hat{X}_3]^2=\Expt[X'-\Expt(X'|U_3)]^2=\frac{d_1\sigma_3^2}{d_1+\sigma_3^2}.
\end{eqnarray}
Note that for optimal distortion quadruples, $d_2\leq d^*_1$ and $d_3\leq d^*_1$ should be chosen, i.e., $\hat{d}_2=d_2$ and $\hat{d}_3=d_3$. 

Notice also
\begin{eqnarray}
I(X;U_2|U_1)&=&I(X';X'+N_2|U_1)\stackrel{(a)}{=}I(X';X'+N_2)\nonumber\\
&=&\frac{1}{2}\log\frac{d_1+\sigma_2^2}{\sigma_2^2},
\end{eqnarray}
where (a) is true because $U_1$ is independent of $X'$ and $N_2$. Choose $\sigma_2^2$ such that
\begin{eqnarray}
R_2\geq \frac{1}{2}\log\frac{d_1+\sigma_2^2}{\sigma_2^2},
\end{eqnarray}
which can always be done because the function is continuous. Similarly choose $\sigma_3^2$ such that 
\begin{eqnarray}
R_3\geq\frac{1}{2}\log\frac{d_1+\sigma_3^2}{\sigma_3^2}.
\end{eqnarray}

From (\ref{eqn:ratesum}), the sum rate satisfying the following bound is achievable
\begin{eqnarray*}
&&R_1+R_2+R_3\\
&\geq& I(X;U_1)+I(X;U_2U_3|U_1)+I(U_2;U_3|U_1)\\
&=&R_1+I(X';U_2U_3|U_1)+I(X'+N_2;X'+N_3|U_1)\\
&=&R_1+I(X';U_2U_3)+I(U_2;U_3)\\
&=&R_1-h(N_2N_3)+h(U_2)+h(U_3),
\end{eqnarray*}
which gives
\begin{eqnarray}
R_2+R_3\geq \frac{1}{2}\log \frac{(d_1+\sigma_2^2)(d_1+\sigma_3^2)}{(1-\rho^2)\sigma_2^2\sigma_3^2}=\frac{1}{2}\log\frac{d_1^2}{d_2d_3(1-\rho^2)}.
\end{eqnarray}
When $\Pi\geq\Delta$, we may choose 
\begin{eqnarray}
\rho = -\sqrt{1-\frac{d_1^2\exp[-2(R_2+R_3)]}{d_2d_3}}.
\end{eqnarray}
Then 
\begin{eqnarray*}
I(X;U_4|U_1U_2U_3)&=&I(X';U_4|U_2U_3)=h(U_4|U_2U_3)-h(N_4)\nonumber\\
&=&\frac{1}{2}\log \frac{\sigma_4^2+d_4^*}{\sigma_4^2},
\end{eqnarray*}
where 
\begin{eqnarray}
d_4^*&\stackrel{\Delta}{=}&\Expt(X'-E(X'|U_2U_3))^2\nonumber\\
&=&\frac{d_1\sigma_2^2\sigma_3^2(1-\rho^2)}{d_1\sigma_2^2\sigma_3^2(1-\rho^2)+d_1(\sigma_2^2+\sigma_3^2)-2\rho d_1\sigma_2\sigma_3}.
\end{eqnarray}
Choose $\sigma_4^2$ such that 
\begin{eqnarray}
R_4=\frac{1}{2}\log \frac{\sigma_4^2+d_4^*}{\sigma_4^2}. 
\end{eqnarray}
We further have (and after some simplification)
\begin{eqnarray}
d_4&=&\Expt(X'-\Expt(X'|U_2U_3U_4))^2\\
&=&\frac{\exp(-2R_4)d_1\sigma_2^2\sigma_3^2(1-\rho^2)}{d_1\sigma_2^2\sigma_3^2(1-\rho^2)+d_1(\sigma_2^2+\sigma_3^2)-2\rho d_1\sigma_2\sigma_3}\\
&=&\frac{\sigma_x^2\exp[-2(R_1+R_2+R_3+R_4)]}{1-(\sqrt{\Pi}-\sqrt{\Delta})^2}.
\label{eqn:normaldelta}
\end{eqnarray}
Thus if $\Pi\geq\Delta$, i.e., $d_1+d_1\exp[-2(R_2+R_3)]\geq d_2+d_3$, then this achievable region matches the outer bounds. 

When $\Pi<\Delta$, i.e., $d_1+d_1\exp[-2(R_2+R_3)]< d_2+d_3$, we can find some $d_2'\leq d_2$ and $d_3'\leq d_3$, where at least one of the inequalities is strict, such that 
$d_1+d_1\exp[-2(R_2+R_3)]=d_2'+d_3'$. To see this, let us first consider choosing $d'_m\in [d_12^{-2R_m},d_m]$, $\ m=2,3$. By a continuity argument, $d'_2+d'_3$ such chosen can take any real value in $[d_1\exp(-2R_2)+d_1\exp(-2R_3),d_2+d_3]$. It is obvious that $d_1[1+\exp(-2(R_2+R_3))]\in [d_1\exp(-2R_2)+d_1\exp(-2R_3),d_2+d_3]$, and it follows there always exists at least a pair of $(d'_2,d'_3)$ such that $d'_2+d'_3=d_1+d_1\exp[-2(R_2+R_3)]$, i.e., $\Pi'=\Delta'$. Thus we can conclude the distortion quadruple $(d_1,d'_2,d'_3,\exp(-2(R_1+R_2+R_3+R_4)))$ is achievable by (\ref{eqn:normaldelta}), which implies that the quadruple $(d_1,d_2,d_3,\exp(-2(R_1+R_2+R_3+R_4)))$ is achievable\footnote{This degenerate region was not treated by Ozarow in \cite{Ozarow:80}, and it sometimes causes certain confusion.}. Therefore for both the case $\Pi\geq\Delta$ and $\Pi<\Delta$ the achievable region indeed matches the outer bounds. 
\end{proof}

\subsection{Fixed channel configuration and the performance loss}
\label{subsec:fixed}

One case of interest is that the good channel and the bad channel used to transmit are fixed as $R_1/R_2=R_3/R_4$, and we will consider the performance loss in this case. Suppose $R_2=\alpha R_1$ and $R_4=\alpha R_3$. 

Consider the cases where only the first layer performance or only the second layer performance is in consideration. For the former case, i.e., the first layer, user 1 has description of rate $R_1$ while user 2 has joint description of rate $(1+\alpha)R_1$, which results in minimum distortions $\sigma_x^2\exp(-2R_1)$ and $\sigma_x^2\exp[-2(1+\alpha)R_1]$. For the latter case, i.e., optimized only for the second layer, user 1 has description of rate $R_1+R_3$ while user 2 has joint description of rate $(1+\alpha)(R_1+R_3)$, which results in minimum distortions $\sigma_x^2\exp[-2(R_1+R_3)]$ and $\sigma_x^2\exp[-2(1+\alpha)(R_1+R_3)]$.

Now for an MSR system to achieve the same minimum second layer distortions as if only the second layer is in consideration. Then by the result from the previous section, we see that $\Pi\leq \Delta$, which gives 
\begin{eqnarray*}
d_2+\sigma_x^2\exp[-2(R_1+R_3)]\geq d_1^*[1+\exp[-2(R_2+R_3)]],
\end{eqnarray*}
which further gives
\begin{eqnarray}
d_2\geq d_1^*[1+\exp(-2(\alpha R_1+R_3))-\exp(-2R_3)].
\end{eqnarray}
For a single layer system optimized for the first layer, distortion $d_1^*=\sigma_x^2\exp(-2R_1)$ and $d_2^*=\sigma_x^2\exp[-2(1+\alpha)R_1]$ are achievable, thus the loss on $d_2$ can be as large as
\begin{eqnarray*}
\frac{d_2}{d_2^*}&=&\frac{[1+\exp(-2(\alpha R_1+R_3))-\exp(-2R_3)]}{\exp(-2\alpha R_1)}\\
&=&\exp(2\alpha R_1)+\exp(-2R_3)-\exp[2(\alpha R_1-R_3)].
\end{eqnarray*}
Thus we see that as $R_1\rightarrow \infty$, \textbf{the performance loss compared to a single layer system can be unbounded}. However, the distortion $d_1$ is not jeopardized by the progressive encoding requirement. In other words $(d_1,d_3,d_4)$ can be matched to an optimal coding system with coding rate $(R_1,R_1+R_3,(1+\alpha)(R_1+R_3))$, with the distortion $d_2$ being quite large. If $d_2$ is of little importance, then such a system can be utilized; otherwise, the performance loss needed to improve $d_2$ is large, and can hardly be compensated by the added functionality. 

\subsection{MD coding vs. Wyner-Ziv coding}

The coding approach proposed in \cite{Pradhanramchandran:02} is based on the Wyner-Ziv (WZ) coding, which treats the reconstruction $\hat{X}_1$ and $\hat{X}_2$ as side informations at the decoder. In this section we compare the performance by the WZ-based coding approach with that by the MD-based coding approach. 

To compare the two coding schemes, fix $d_1=d_1^*=\sigma_x^2\exp(-2R_1)$ and $d_2=d_2^*=\sigma_x^2\exp[-2(R_1+R_2)]$. For the WZ-based approach, since the $\phi_3$ and $\phi_4$ are  successive refinement by definition, the WZ-based coding is in fact the successive Wyner-Ziv problem with degraded side information at the decoder considered by Steinberg and Merhav in \cite{SteinbergMerhav:04}. Though $\hat{X}_1$ and $\hat{X}_2$ are not necessarily  physically degraded, as pointed out in \cite{SteinbergMerhav:04}, the achievable region is only dependent on the pairwise distribution between the source and the side information, thus statistical degradedness 
 and physical degradedness have no essential difference. It is known that the Gaussian source and side informations can always be taken as statistically degraded, and thus the general result in \cite{SteinbergMerhav:04} can be readily used. The rate-distortion region for the Gaussian source was given explicitly in \cite{TianDiggavi:072}, and can be (modified accordingly and) written as follows. Choose $\sigma_1^2$ and $\sigma_2^2$ such that
\begin{eqnarray}
d_1^*=\frac{\sigma_x^2(\sigma_1^2+\sigma_2^2)}{\sigma_x^2+\sigma_1^2+\sigma_2^2},\quad d_2^*=\frac{\sigma_x^2\sigma_2^2}{\sigma_x^2+\sigma_2^2}
\end{eqnarray}
and define $\gamma\stackrel{\Delta}{=} \frac{\sigma_2^2}{\sigma_1^2+\sigma_2^2}$,
then the achievable distortions using WZ-based coding are given by 
\begin{eqnarray}
d_3'&\geq& \exp(-2R_3)d_3^*=\sigma_x^2\exp[-2(R_1+R_3)]\\
d_4'&\geq& \exp[-2(R_3+R_4)]\nonumber\\
&&\times
\frac{\sigma_x^2\sigma_1^2\sigma_2^2}{(\sigma_x^2+\sigma_1^2+\sigma_2^2)((1-\gamma)^2\min(d_3',d_1^*)+\gamma\sigma_1^2)}.
\end{eqnarray}

On the other hand, the MD-based coding approach can achieve 
\begin{eqnarray}
d_3&\geq& \sigma_x^2\exp[-2(R_1+R_3)],\\
d_4&\geq& \frac{\sigma_x^2\exp[-2(R_1+R_2+R_3+R_4)]}{1-(|\sqrt{\Pi}-\sqrt{\Delta}|^+)^2},
\end{eqnarray}
where we have for this special case
\begin{eqnarray}
\Pi=(1-\frac{d_2^*}{d_1^*})(1-\frac{d_3}{d_1^*})=[1-\exp(-2R_2)](1-\frac{d_3}{d_1^*}),
\end{eqnarray}
as well as
\begin{eqnarray}
\Delta&=&\frac{d_2^*d_3}{d_1^{*2}}-\exp[-2(R_2+R_3)]\nonumber\\
&=&\exp(-2R_2)[\frac{d_3}{d_1^{*}}-\exp(-2R_3)].
\end{eqnarray}

\begin{figure}[tb]
  \centering 
\includegraphics[scale=0.45]{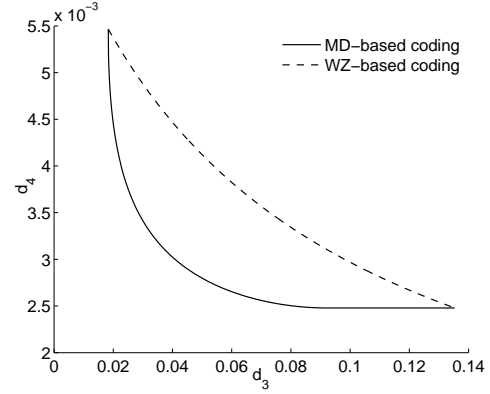}
\caption{Comparison of the distortion regions $(d_3,d_4)$ of the two coding approaches at $R_1=R_3=1.0$ nat and $R_2=R_4=0.5$ nat with fixed $d_1=d_1^*$ and $d_2=d_2^*$ \label{fig:WZMD}. The range of $d_3$ is $[\sigma_x^2\exp[-2(R_1+R_3)],d_1^*]$.}
\end{figure}

A set of typical tradeoff curves of $(d_3,d_4)$ for the WZ-based approach and MD-based approach are given in Fig. \ref{fig:WZMD} for fixed rates $(R_1,R_2,R_3,R_4)$, and we can see the gap is non-zero. As such, the WZ-based approach is suboptimal except two extreme operating points: one point is when $d_3=d_1^*$ and the other one is $d_3=\sigma_x^2\exp[-2(R_1+R_3)]$, which correspond to description $\phi_3$ is either completely useless for decoder 3, or maximally useful for decoder 3, respectively. This in fact illustrates the role of side informations at only the decoders or at both the encoder and the decoders are quite different: it is known that for the Gaussian source there is no loss between the cases when a {\em single} Gaussian side information is available at both the encoder and the decoder, or at the decoder only, however when there are multiple side informations at different decoders, they are no longer equivalent. This observation perhaps was firstly made explicit in \cite{Perron:06}. Though the Wyner-Ziv coding based approach is sub-optimal in this problem, it does have certain advantage, particularly when the second round of transmission is not encoded together with the first round descriptions, but is made possible when certain network resource becomes available after an initial transmission. 

\begin{figure}[tb]
  \centering 
\includegraphics[scale=0.3]{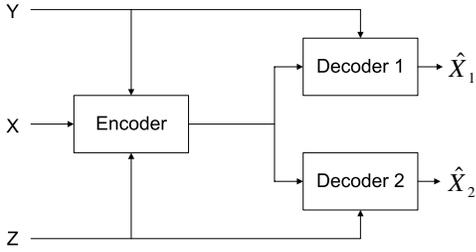}
\caption{The system diagram for the EDSI problem\label{fig:EDSI}.}
\end{figure}

The MSR problem in fact has a more subtle connection with the encoder/decoder side information (EDSI) problem considered in \cite{Perron:06} (see also \cite{Wolf:04}), which is depicted in Fig. \ref{fig:EDSI}. The connection is through one particular special case for the EDSI problem, when the source and side informations are physically degraded as $X\leftrightarrow Y \leftrightarrow Z$, and in the Gaussian case we may write without loss of generality $Y=X+N_1$ and $Z=Y+N_2$ where $N_1$ and $N_2$ are independent Gaussian noise. 
Now if we take $\hat{X}_1$ and $\hat{X}_2$ in MSR as the side informations $Z$ and $Y$, respectively, and let $R_4=0$, then MSR can be considered as a relaxed version of the EDSI problem, because in MSR the codeword $\hat{X}_1$ and $\hat{X}_2$ do not have to be generated by any marginal distribution as specified in the EDSI problem, but we can indeed choose them to have such structure with single-letter distribution $P_{ZY}$; furthermore, in the MSR problem, decoder one always has $\hat{X}_1$ (corresponding to $Z$), rather than only $\hat{X}_2$ (corresponding to the better side information $Y$). As such, if in the MSR system we set $R_1=I(X;Z)$ and $R_2=I(X;Y|Z)$, $D_1=\Expt[X-\Expt(X|Z)]^2$ and $D_2=\Expt[X-\Expt(X|Y)]^2$, an outer bound for the EDSI problem can be found; it is an outer bound since we can use $P_{XZY}$ as $P_{XU_1U_2}$ in the minimization for the MSR problem, but the chosen $(R_1,R_2,D_1,D_2)$ also allows for other choices of random variables. As shown in \cite{Perron:06}, this outer bound is indeed achievable by using a hybrid conditioning/binning scheme \footnote{The (same) outer bound for this case given in \cite{Perron:06} was derived by applying the conditional version of the results of \cite{Kaspi:94}, which is indeed closely related to the MD problem.}. 

\section{A Variation of the MD Problem: the MDCR Problem}
\label{sec:variation}

As aforementioned, when $R_1=0$, the problem being considered reduces to the MD problem with central refinement (MDCR), and the six-tuple of rates and distortions are to be characterized. Again we focus on the Gaussian case; we will continue to use the notations $(R_2,R_3,R_4)$ and descriptions $\phi_2,\phi_3,\phi_4$ and assuming $R_1=0$ and no description $\phi_1$ exists.

El Gamal and Cover constructed an MD scheme for general sources based on the MDCR method in \cite{Gam:82}. More precisely, the description $\phi_4$ of rate $R_4$ is split and combined into the existing two descriptions $\phi_2$ and $\phi_3$, and the resulting two descriptions are of rates $R_2'=R_2+\beta R_4$ and $R_3'=R_3+(1-\beta)R_4$ for some $0\leq\beta\leq 1$.   
It is known that in the Gaussian MD problem, there is no need for the central refinement coding to achieve the complete distortion region; i.e., $R_4=0$ is sufficient to achieve the complete distortion region given the two description rates $R_2'$ and $R_3'$. 
A natural question to ask is whether it is possible to construct an optimal Gaussian MD system using an MDCR system with nonzero rate $R_4$.

The answer to the above question in fact negative, which is implied by Theorem \ref{theorem:Gaussian}. To see this, assume the distortion $\tilde{d}_2$ and $\tilde{d}_3$ in both an MDCR-based system and an optimal MD system, such that $\tilde{d}_2<d^*_1$ and $\tilde{d}_3<d^*_1$. By Theorem \ref{theorem:Gaussian}, we see that for an MDCR system with non-zero $R_4$
\begin{eqnarray}
d_4\geq\frac{\sigma_x^2\exp[-2(R_2+R_3+R_4)]}{1-(\sqrt{\Pi}-\sqrt{\Delta})^2}
\end{eqnarray}
where 
\begin{eqnarray}
\Delta=\frac{\tilde{d}_2\tilde{d}_3}{\sigma_x^4}-\exp[-2(R_2+R_3)], \quad
\Pi=(1-\frac{\tilde{d}_2}{\sigma_x^2})(1-\frac{\tilde{d}_3}{\sigma_x^2}).
\end{eqnarray}
For an optimal MD system, the distortion resulting from the joint description can be 
\begin{eqnarray}
d_4'\geq\frac{\sigma_x^2\exp[-2(R_2'+R_3')]}{1-(\sqrt{\Pi'}-\sqrt{\Delta'})^2}
\end{eqnarray}
where 
\begin{eqnarray}
\Delta'=\frac{\tilde{d}_2\tilde{d}_3}{\sigma_x^4}-\exp[-2(R_2'+R_3')],\,
\Pi'=(1-\frac{\tilde{d}_2}{\sigma_x^2})(1-\frac{\tilde{d}_3}{\sigma_x^2}).
\end{eqnarray}
To keep the rates of the two system equal, we have $R_2'=R_2+\beta R_4$ and $R_3'=R_3+(1-\beta)R_4$ for some $0\leq\beta\leq 1$; further assume $\tilde{d}_2$ and $\tilde{d}_3$ are chosen such that $\Pi'\geq\Delta'$, i.e., the MD system does not operate in the degenerate region. It is now seen that the MDCR approach with non-zero $R_4$ is suboptimal because $d_4>d_4'$, due to the fact that $\Delta'>\Delta$ and $\Pi'=\Pi$. This is a stronger result than the known one that it is sufficient for $R_4=0$ to achieve optimality: \textbf{it is in fact necessary for $R_4$ to be zero in order to be optimal in the Gaussian MD case in general}. 

This result suggests any system based on the MDCR approach when the refinement rate is not zero is not optimal for the Gaussian source: one such example is the system constructed with dithered lattice quantizers in \cite{Zam:02}. 

\subsection{The high-rate asymptotics for balanced descriptions}
We consider balanced MDs in this subsection, and further assume $\sigma_x^2=1$. Suppose in an MD system, the rate of two descriptions are at equally high rate of $R'$ each, and the side distortions are both $d_2'=d_3'$. It can be shown that if the side distortion is of the form $d_2^\prime=b2^{-2(1-\eta) R^\prime}$, where $0\leq\eta<1$ and $b\geq1$, the central distortion of an MD system can asymptotically (at low distortion) achieve 
\begin{equation}
d_4^\prime\geq\left\{\begin{array}{lr}
              \textstyle{2^{-2R^\prime}/2(b+\sqrt{b^2-1})}& \eta=0;\\
              \textstyle{2^{-2R^\prime(1+\eta)}/4b}  & 0<\eta<1.\\              
              \end{array}\right.     
\label{equ:prod2}         
\end{equation}
Notice the condition $0<\eta<1$ in fact corresponds to the condition that $1\gg{d_2^\prime}$ and $d_2^\prime\gg{d_4^\prime}$ at high rate. In this case, the central and side distortions' product remains bounded by a constant at fixed rate, which is $d_4^\prime d_2^\prime\geq\frac{2^{-4R^\prime}}{4}$, independent of the tradeoff between them. This product has been used as the information theoretical bound to measure the efficiency of quantization methods \cite{Cha:041,Vai:01}. Below, the performance of the optimal MD system is compared with that of an MDCR-based system in this high-rate and high-refinement-rate case. 

For an MDCR-based MD system, $R_4$ is allocated to the refinement stage, and thus each of the first stage descriptions is of rate $R^\prime-R_4/2$. Keeping the side distortion of this system $d_2=d_2^\prime=b2^{-2(1-\eta)R^\prime}$ for an easier comparison, consider the case $1>\eta>0$, and let $R_4=2\eta_1 R^\prime$, where $1-\eta_1$ is the ratio between $R_2$ and $R_2'$.
Then it can be shown (through some algebra) that using the MDCR approach, we can achieve
\begin{equation}
d_4\geq\left\{\begin{array}{lr}
              \textstyle{2^{-2R^\prime(1+\eta)}/2(b+\sqrt{b^2-1})}& \eta_1=\eta;\\
              \textstyle{2^{-2R^\prime(1+\eta)}/4b}  & 0\leq\eta_1<\eta.\\              
              \end{array}\right.     
\end{equation}
This implies that if the first stage has sufficient excess marginal rate, i.e., $\eta_1<\eta$, then the performance loss from the optimal MD system by the MDCR approach with non-zero $R_4$, in terms of the distortion product, is asymptotically zero in the range of $1\gg{d_2}$ and $d_2\gg{d_4}$. However, as the rate allocated to the refinement stage increases, the excess marginal rate in the first stage decreases. When $\eta_1=\eta$, the performance loss is a factor of $\frac{2b}{(b+\sqrt{b^2-1})}$. If the first stage is without excess marginal rate, which means $\eta_1=\eta$ and $b=1$, then the loss is a factor of $2$ comparing to the MD system without taking such an MDCR approach. 

This discussion suggests that the MDCR approach is appealing for the high-rate case, if $1\gg{d_2}$ and $d_2\gg{d_4}$ is the desired operating range. However, the first stage should reserve sufficient excess marginal rate in order to avoid the performance loss. Taking the MD system in \cite{Zam:02} as an example, using certain sub-optimal lattices for $\phi_2$ and $\phi_3$ is potentially able to achieve (asymptotic) optimal performance, but using two good lattices as $\phi_2$ and $\phi_3$ will not be, because the excess marginal rate is diminishing as the dimension increases. 

\section{Conclusion}
\label{sec:conclusion}
We considered the problem of multiuser successive refinement. An achievable region is provided, which is shown to be tight for the Gaussian source under MSE measure. It is shown that different from the single user case, the MSR coding necessitates performance loss, which can be unbounded. The results rely on the recognition that a multiple description system is embedded inside the MSR system. The MSR system also includes a variation of the MD system, namely the MDCR problem. This problem is treated with some depth, which reveals some interesting implications in designing the MD coding system.

For the general problem with an arbitrary $K>2$ rounds of transmission, or $K>2$ users, an achievable region can be derived using the technique developed in \cite{Venkataramani:03} and \cite{Pradhan:04}. However, even for the Gaussian case, the problem is highly intractable, and a complete characterization appears difficult. Given the results in the current work, we expect the loss of performance for the general case to be more severe than the $K=2$ case.


\bibliographystyle{IEEEbib}

\begin{thebibliography}{8}

\bibitem{Pradhanramchandran:02}
S.~Pradhan and K.~Ramchandran,
\newblock ``Multiuser successive refinement,''
\newblock in {\em Conference on Information Sciences and Systems}, Mar. 2002.

\bibitem{Koshelev:80}
V.~N. Koshelev,
\newblock ``Hierarchical coding of discrete sources,''
\newblock {\em Probl. Pered. Inform.}, vol. 16, no. 3, pp. 31--49, 1980.

\bibitem{EquitzCover:91}
W.~H.~R. Equitz and T.~M. Cover,
\newblock ``Successive refinement of information,''
\newblock {\em IEEE Trans. Information Theory}, vol. 37, no. 2, pp. 269--275,
  Mar. 1991.

\bibitem{Rimoldi:94}
B.~Rimoldi,
\newblock ``Successive refinement of information: Characterization of
  achievable rates,''
\newblock {\em IEEE Trans. Information Theory}, vol. 40, no. 1, pp. 253--259,
  Jan. 1994.

\bibitem{Lastras:01}
L.~Lastras and T.~Berger,
\newblock ``All sources are nearly successively refinable,''
\newblock {\em IEEE Trans. Information Theory}, vol. 47, no. 3, pp. 918--926,
  Mar. 2001.

\bibitem{Ozarow:80}
L.~Ozarow,
\newblock ``On a source-coding problem with two channels and three receivers,''
\newblock {\em Bell Syst. Tech. Journal}, vol. 59, pp. 1909--1921, Dec. 1980.

\bibitem{Gam:82}
A.~El Gamal and T.~M. Cover,
\newblock ``Achievable rates for multiple descriptions,''
\newblock {\em IEEE Trans. Information Theory}, vol. 28, no. 6, pp. 851--857,
  Nov. 1982.

\bibitem{Wynerziv:76}
A.~D. Wyner and J.~Ziv,
\newblock ``The rate-distortion function for source coding with side
  information at the decoder,''
\newblock {\em IEEE Trans. Information Theory}, vol. 22, no. 1, pp. 1--10, Jan.
  1976.

\bibitem{Perron:06}
E.~Perron, S.~Diggavi, and E.~Telatar,
\newblock ``On the role of encoder side-information in source coding for
  multiple decoders,''
\newblock in {\em Proc. 2006 International Symposium on Information Theory},  
  Seattle, WA, Jul. 2006, pp. 331--335.

\bibitem{TianChen:05}
C.~Tian, J.~Chen, S.~Hemami, and T.~Berger,
\newblock ``Multiple descriptions with central refinement,''
\newblock in {\em Conference on Information Sciences and Systems}, Mar. 2005.

\bibitem{FengEffros:031}
H.~Feng and M.~Effros,
\newblock ``On the achievable region for multiple description source codes on
  {Gaussian} sources,''
\newblock in {\em Proc. IEEE Symposium Information Theory}, Jun.-Jul 2003, p. 195.

\bibitem{FengEffros:07}
H.~Feng and M.~Effros,
\newblock ``Nearly separability of optimal multiple description source codes,''
\newblock {\em IEEE Trans. Information Theory}, submitted for publication.


\bibitem{Gastpar:03}
M.~Gastpar, B.~Rimoldi, and M.~Vetterli,
\newblock ``To code, or not to code: lossy source-channel communication
  revisted,''
\newblock {\em IEEE Trans. Information Theory}, vol. 49, no. 5, pp. 1147--1158,
  May. 2003.

\bibitem{SteinbergMerhav:06}
Y.~Steinberg and N.~Merhav,
\newblock ``On hierarchical joint source-channel coding with degraded side
  information,''
\newblock {\em IEEE Trans. Information Theory}, vol. 52, no. 3, pp. 886--903,
  Mar. 2006.

\bibitem{TianDiggavi:072}
C.~Tian and S.~Diggavi,
\newblock ``On multistage successive refinement for wyner-ziv source coding
  with degraded side information,''
\newblock {\em IEEE Trans. Information Theory}, vol. 53, no. 8, pp. 2946--2960 ,
  Aug. 2007.

\bibitem{Goyal:011}
V.~K. Goyal,
\newblock ``Multiple description coding: {C}ompression meets the network,''
\newblock {\em IEEE Signal Processing Magazine}, vol. 47, no. 6, pp. 2199 --
  2224, Sep. 2001.

\bibitem{Lastras:06}
L.~A. Lastras and V.~Castelli,
\newblock ``Near sufficiency of random coding for two descriptions,''
\newblock {\em IEEE Trans. Information Theory}, vol. 52, no. 2, pp. 618--695,
  Feb. 2006.

\bibitem{ZhangBerger:87}
Z.~Zhang and T.~Berger,
\newblock ``New results in binary multiple descriptions,''
\newblock {\em IEEE Trans. Information Theory}, vol. 33, no. 4, pp. 502--521,
  Jul. 1987.

\bibitem{WangPramod:05}
H.~Wang and P.~Viswanath,
\newblock ``Vector {Gaussian} multiple description with individual and central
  receivers,''
\newblock {\em IEEE Trans. Information Theory}, vol. 53, no. 6, pp. 2133--2153,
  Jun. 2007.

\bibitem{Sgarro:77}
A.~Sgarro,
\newblock ``Source coding with side information at several decoders,''
\newblock {\em IEEE Trans. Information Theory}, vol. 23, no. 2, pp. 179--182,
  Mar. 1977.

\bibitem{Pradhan:04}
S.~S. Pradhan, R.~Puri, and K.~Ramchandran,
\newblock ``n-channel symmetric multiple descriptions - {P}art {I}: (n, k)
  source-channel erasure codes,''
\newblock {\em IEEE Trans. Information Theory}, vol. 50, pp. 47--61, Jan. 2004.

\bibitem{DiggaviVinay:04}
S.~Diggavi and V.~Vaishampayan,
\newblock ``On multiple description source coding with decoder side
  information,''
\newblock in {\em Proc. IEEE Information Theory Workshop}, San Antonio, TX,
  Oct. 2004.

\bibitem{Effros:99}
M.~Effros,
\newblock ``Distortion-rate bounds for fixed- and variable-rate multiresolution
  source codes,''
\newblock {\em IEEE Trans. Information Theory}, vol. 45, no. 6, pp. 1887--1910,
  Sep. 1999.

\bibitem{Gallagerbook}
R.~G. Gallager,
\newblock {\em Information theory and reliable communication},
\newblock New York: John Wiley, 1968.

\bibitem{Wyner:78}
A.~D. Wyner,
\newblock ``The rate-distortion function for source coding with side
  information at the decoder {II}: general sources,''
\newblock {\em Inform. contr.}, vol. 38, pp. 60--80, 1978.

\bibitem{ChenTian:05}
J.~Chen, C.~Tian, S.~Hemami, and T.~Berger,
\newblock ``Multiple description quantization via {Gram-Schmidt}
  orthogonalization,''
\newblock {\em IEEE Trans. Information Theory}, vol. 52, no. 12, pp.
  5197--5217, Dec. 2006.

\bibitem{CoverThomas}
T.~M.~Cover and J.~A.~Thomas,
\newblock {\em Elements of information theory},
\newblock New York: Wiley, 1991.

\bibitem{SteinbergMerhav:04}
Y.~Steinberg and N.~Merhav,
\newblock ``On successive refinement for the {Wyner-Ziv} problem,''
\newblock {\em IEEE Trans. Information Theory}, vol. 50, no. 8, pp. 1636--1654,
  Aug. 2004.

\bibitem{Wolf:04}
J.~K. Wolf,
\newblock ``Source coding for a noiseless broadcast channel,''
\newblock in {\em Conference on Information Sciences and Systems}, Mar. 2004.

\bibitem{Kaspi:94}
A.~Kaspi,
\newblock ``Rate-distortion when side-information may be present at the
  decoder,''
\newblock {\em IEEE Trans. Information Theory}, vol. 40, no. 6, pp. 2031--2034,
  Nov. 1994.

\bibitem{Zam:02}
Y.~Frank-Dayan and R.~Zamir,
\newblock ``Dithered lattice-based quantizers for multiple descriptions,''
\newblock {\em IEEE Trans. Information Theory}, vol. 48, no. 1, pp. 192--204,
  Jan. 2002.

\bibitem{Cha:041}
C.~Tian and S.~S.~Hemami,
\newblock ``Universal multiple description scalar quantizer: analysis and
  design,''
\newblock {\em IEEE Trans. Information Theory}, vol. 50, no. 9, pp. 2089--2102,
  Sep. 2004.

\bibitem{Vai:01}
V.~A. Vaishampayan, N.~J.~A. Sloane, and S.~D. Servetto,
\newblock ``Multiple-description vector quantization with lattice codebooks:
  design and analysis,''
\newblock {\em IEEE Trans. Information Theory}, vol. 47, no. 5, pp. 1718--1734,
  Jul. 2001.

\bibitem{Venkataramani:03}
R.~Venkataramani, G.~Kramer, and V.~K.~Goyal,
\newblock ``Multiple description coding with many channels,''
\newblock {\em IEEE Trans. Information Theory}, vol. 49, no. 9, pp. 2106--2114,
  Sep. 2003.

\end{thebibliography}

\end{document}